\documentclass[12pt]{iopart}

\usepackage{amssymb}
\usepackage[compress]{cite}
\usepackage{txfonts}
\usepackage{iopams}
\usepackage{graphics}
\usepackage{color} 
\usepackage{CJK}

\newcommand{\beq}{\begin{equation}}
\newcommand{\eeq}{\end{equation}}

\newcommand{\beqa}{\begin{eqnarray}}
\newcommand{\eeqa}{\end{eqnarray}}
\newcommand{\beqas}{\begin{eqnarray*}}
\newcommand{\eeqas}{\end{eqnarray*}}
\newcommand{\ra}{\rangle}
\newcommand{\la}{\langle}

\begin{document}

\title[Optimal shortcuts for atomic transport in anharmonic traps]{Optimal shortcuts for atomic transport in anharmonic traps}

\author{Qi Zhang$^{1}$, J. G. Muga$^{1,2}$, D. Gu\'{e}ry-Odelin$^{3}$, and Xi Chen$^{1,4}$}

\address{$^{1}$ Department of Physics, Shanghai University, 200444 Shanghai, P. R. China \\
         $^{2}$ Departamento de Qu\'{\i}mica-F\'{\i}sica, UPV-EHU, Apdo 644, 48080 Bilbao, Spain \\
         $^{3}$ Laboratoire Collisions Agr\'egats R\'eactivit\'e, CNRS UMR 5589, IRSAMC, Universit\'e Paul Sabatier, 118 Route de Narbonne, 31062 Toulouse CEDEX 4, France \\
         $^{4}$ State Key Laboratory of Precision Spectroscopy, East China Normal University, Shanghai 200062, P. R. China
         }
\ead{xchen@shu.edu.cn}

\vspace{10pt}
\begin{indented}
\item[]February 2016
\end{indented}

\begin{abstract}
We design optimal trajectories to transport cold atoms in anharmonic traps,
combining invariant-based inverse engineering, perturbation theory, and optimal control theory.
The anharmonic perturbation energy is minimized constraining the
maximally allowed relative displacement between the trap center and the atom.
\end{abstract}

%
%
%
%
%

\section{Introduction}
A major goal of modern physics is to achieve a thorough control of the motional and internal state of the atom
preserving quantum coherence and avoiding undesired excitations.
In particular, many experiments and proposals to develop quantum technologies
require to shuttle cold neutral atoms or ions by moving the confining trap,
leaving them at rest and unexcited at the destination
site
\cite{HanschNature2001,HanschPRL2001,Ketterle2002,Leibfried,ions,Bowler,Walther}.
Several approaches with small (ideally negligible) final excitation but moderate transient motional excitation,
have been put forward to achieve fast non-adiabatic transport
\cite{David08,Calarco09,TransAPL10,Masuda10,Erik11,Xi12,Erikcond12,Mikel13,Uli14,CampoPRX,Lu14,Mikel14,David14,Qi2015}.
Reducing the transport time with respect to adiabatic times (long times for which even transient excitations are  suppressed)
is of interest to achieve faster operations, e.g., in quantum information processing,
and also to  avoid overheating from fluctuating fields and decoherence.
In particular, the combination of invariant-based inverse engineering and optimal control theory, is a versatile toolbox
for designing optimal transport protocols, according to different physical criteria or operational constraints \cite{Li2010,Xi12,Erikcond12}.
Furthermore, fast transport can be further optimized with respect to spring-constant errors \cite{David14}, spring-constant (colored) noise, and position fluctuations
\cite{Lu14}.

Different transport protocols have been designed
for harmonic traps but of course actual confining traps such as magnetic quadrupole potentials \cite{TransAPL10}, gravitomagnetric potentials \cite{Ricci}, Penning-trap potentials \cite{AlonsoNJP}, and optical dipole traps \cite{OptTrap}, are anharmonic.
The anharmonic terms limit the validity of harmonic approximations and thus the possible process speeds \cite{Chenenergy10}.
Their effect has been studied for three-dimensional optical traps \cite{Erik3DPRA}, perturbatively for condensates \cite{Erikcond12},
and classically in \cite{Qi2015}. Ref. \cite{Mikel13} analyzed as well the coupling between center-of-mass  and relative motions of two ions
due to anharmonicity. It is known that a force proportional to the acceleration of the trap
exactly compensates for the inertial force in the moving trap frame, even for anharmonic potentials,  avoiding any excitation
\cite{Masuda10,Erik11}.
It has been pointed out, however, that this force may be difficult to implement in some systems,
such as chains of ions of different mass \cite{Erik11,Mikel13}, or due to practical limitations in the strength of the applicable force \cite{Mikel13}, so that alternative approaches are worth pursuing.
A missing piece in the existing studies was an optimal control theory solution, similar to the ones found for expansions
of anharmonic traps \cite{LuPRA2014}. The aim of this paper is to fill that gap. Even if the optimal protocols may be difficult to implement, typically because of discontinuities
or jumps in the control parameters, they set a useful reference and bounds that limit what can be achieved  with smoother, suboptimal versions.

%
%
%
%
%
\section{Model, Dynamical Invariants, and Perturbation Theory}
\label{sec2}
\subsection{Model}
\begin{figure}[tbp]
\begin{center}
\scalebox{0.8}[0.8]{\includegraphics{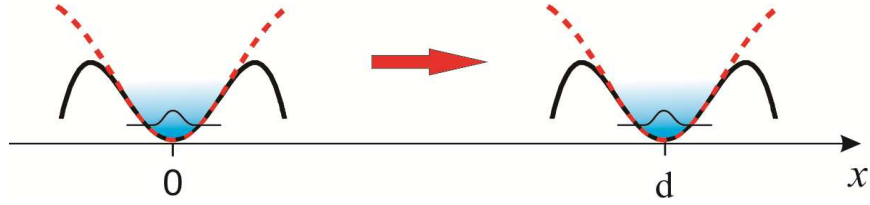}}
\caption{\label{fig1} (Color online) Schematic diagram of atomic transport in an effective one-dimensional Gaussian trap (dashed red line), approximated as a harmonic trap plus anharmonic quartic term (solid black line) from $x=0$ to $x=d$.}
\end{center}
\end{figure}
We shall consider the following Hamiltonian model for a single particle of mass $m$ moving in one dimension (with coordinate $x$) in a moving, anharmonic potential,  
\beq
\label{H+V}
H(t)= \frac{p^2}{2m}+\frac{1}{2}m\omega_0^2[x-x_0(t)]^2 - \eta [x-x_0(t)]^4,
\eeq
where $p$ is the momentum operator. As a concrete example, we consider the on-axis potential produced by an optical tweezer made of a focussed Gaussian beam \cite{David08,Erik3DPRA}. Its expansion about its minimum (see Fig.~\ref{fig1}) yields $\omega_0=(2V_0/mz_R^2)^{1/2}$ and $\eta=V_0/z_R^4$, where $V_0$ is the depth of the potential, $z_R=\pi w_0^2/\lambda$ the Rayleigh length, $w_0$ the waist of the Gaussian beam, and $\lambda$ its wavelength. In the following we choose parameters close to those of the experimental work of Ref.~\cite{David08}: $d=1 \times 10^{-2}$~m, $\omega_0=2 \pi\times 20 $~Hz, $w_0=50\lambda$, $\lambda=1060$~nm, $m= 1.44269 \times 10^{-25}$~kg, the mass of $^{87}\mbox{Rb}$ atoms.
%

In general, the Hamiltonian (\ref{H+V}) does not belong to the family of Lewis-Leach
potentials compatible with quadratic-in-momentum invariants \cite{Leach}, so invariant-based engineering cannot be applied
directly, as it is done for the purely harmonic trap \cite{Erik11,Xi12,Erikcond12}. One way out is to add a linear term and thus a compensating force.
Some difficulties with this approach will be pointed out in  Sec. \ref{cf}.
A second strategy, which will take the main part of this work, is to work out first the family of shortcuts for a purely harmonic trap,
and then combine perturbation theory and optimal control theory to
design optimally fast atomic transport, by minimizing the contribution of the anharmonicity to the
potential energy.
%
%
%
\subsection{Harmonic potential and invariant}
We first review briefly the invariant-based inverse engineering approach for (one-dimensional) atomic transport in harmonic traps \cite{Erik11,Xi12,Erikcond12}.
Harmonic transport is described by the Hamiltonian
\beq
\label{Hamiltonian}
H_0 (t)= \frac{p^2}{2 m} + \frac{1}{2}m \omega^2_0 [x-x_0(t)]^2.
\eeq
It has the quadratic-in-momentum Lewis-Riesenfeld invariant \cite{LR,Leach}
\beq
\label{inva}
I (t) = \frac{1}{2m}[p-m\dot{x}_c(t)]^2 +\frac{1}{2}m \omega^2_0 [ x-x_c (t)]^2,
\eeq
provided $x_c (t)$ satisfies Newton's equation,
\beq
\label{classical}
\ddot{x}_c+\omega_0^2(x_c-x_0)=0,
\eeq
for a classical trajectory in the moving harmonic potential. To get $[H_0(t), I(t)]=0$, at $t=0$
and $t=t_f$, so that the Hamiltonian and invariant operators share the same eigensates
at the boundary times, as well as $x_0=x_c$ at the boundary times, we impose
\beqa
\label{con0}
x_c(0) &=& \dot{x}_c(0)= \ddot{x}_c(0)=0,
\\
\label{contf}
x_c(t_f)&=& d;\; \dot{x}_c(t_f)= \ddot{x}_c(t_f)=0,
\eeqa
and interpolate $x_c(t)$ in between,
for example, by a simple polynomial ansatz,
\beq
\label{poly}
x_c(t) =10 d (t/t_f)^3 - 15 d (t/t_f)^4 + 6 d (t/t_f)^5.
\eeq
The imposed boundary conditions guarantee that there is no final vibrational excitation when the trap is moved
from  $x_0 (0)=0$ at time $t=0$ to $x_0(t_f)=d$ at $t_f$.
The ``transport modes'' are solutions of the time-dependent Schr\"odinger equation given by
eigenstates of the dynamical invariant $I(t)$ multiplied by the Lewis-Riesenfeld phase factors,
and can be written as \cite{Erik11}
\beqa
\label{psin}
\la x | \psi_n(t) \ra &=& \frac{1}{(2^n n!)^{1/2}} \left( \frac{m \omega_0}{\pi \hbar}\right)^{1/4}
\nonumber\\
&\times&\exp\left[{ -\frac{i}{\hbar}\int_0^t  {\rm d} t' \left(\lambda_n+\frac{m\dot{x}^2_c}{2} \right)}\right]
\nonumber \\ &\times&
 \exp{\left[- \frac{m \omega_0}{2\hbar} (x-x_c)^2\right]} \exp{\left(i \frac{ m\dot{x}_c x}{\hbar}\right)}\nonumber \\
 &\times& H_n \left[\left(\frac{m \omega_0}{\hbar}\right)^{1/2}(x-x_c)  \right],
\eeqa
where $\lambda_n=(n+1/2)\hbar \omega_0$ is real time-independent eigenvalue of the invariant and $H_n$ is a Hermite polynomial.
An arbitrary solution of the time-dependent Schr\"odinger equation
$i \hbar \partial_t\Psi (x,t) = H_0 (t) \Psi (x,t)$, can be written as
$ \Psi(x,t) = \sum_n c_n \psi_n(x,t)$, where
$n=0,1,...$ and $c_n$ are time-independent coefficients.
The instantaneous average
energy for a transport mode can be obtained from (\ref{Hamiltonian})
and (\ref{psin}),
\beqa
\langle \psi_n(t)|H_0(t)|\psi_n(t)\rangle = {\hbar\omega_0}\left(n+1/2\right) + E_{c} + E_{p},
\eeqa
where the first, ``internal'' contribution remains constant for each $n$,
$E_{c} =m \dot{x}_c^2/2$, and $E_p= m\omega_0^2(x_c-x_0)^2/2$
has the form of a potential energy for a classical particle.
The instantaneous average potential energy can be written as
\beq
\label{potener}
\la V(t)\ra=\frac{\hbar\omega_0}{2}\left(n+1/2\right)+E_p,
\eeq
where here $V=m\omega_0^2(x-x_0)^2/2$.
\subsection{Compensating force\label{cf}}
Any moving potential $V[x-x_0(t)]$ can in fact be used for excitation-free transport if a linear term $-mx\ddot{x}_0$ is superimposed to compensate for the
inertial force \cite{Masuda10,Erik11}.
In particular the potential in (\ref{H+V}) has to be substituted by
\beq
V_{c}=\frac{1}{2}m\omega_0^2 (x-x_0)^2+\eta(x-x_0)^4-mx\ddot{x}_0,
\eeq
which may be rewritten as
\beq
V_{c}=\frac{1}{2} m \widetilde{\omega}_0^2(x-\widetilde{x}_0)^2+B(x^3,x^4)+C,
\eeq
in which the new time-dependent frequency is
\beq
\widetilde{\omega}_0=\sqrt{\omega_0^2+\frac{12}{m}\eta x_0^2},
\eeq
and the new center of the harmonic part is
\beq
\label{x0pp}
\widetilde{x}_0=\frac{\frac{1}{2}m\omega_0^2 x_0+\frac{1}{2}m \ddot{x}_0+2\eta x_0^3}{A},
\eeq
where $A = m \widetilde{\omega}^2_0/2$, $B = \eta(x^4-4x^3 x_0)$,
and
$$
C = \eta x_0^4\!+\!\frac{m}{2}\omega_0^2 x_0^2\!-\!\left(\frac{m}{2}\omega_0^2 x_0\!+\!\frac{m}{2} \ddot{x}_0\!+\!2\eta x_0^3\right)^{\!\!2}\!\!/\!A.
$$
$C$ is an irrelevant purely time-dependent term.
For a purely harmonic trap, $\eta=0$, the compensating force simply amounts to shifting the motion of
the original trap,
see (\ref{x0pp}), as $\widetilde{\omega}_0=\omega_0$ in this case.
If $\eta\ne 0$, however, the time-dependent potential is not simply a displaced copy of the original one: the harmonic frequency changes with time,
and a cubic term appears. Implementing the protocol becomes challenging, as a direct realization of the linear term is
limited by experimental constraints, which are more stringent for neutral atoms, for example due to limits on the
magnetic field gradient,  than for trapped ions \cite{Uli14}, where an extra electric field is easy to implement.
This motivates an alternative approach that combines inverse engineering with optimal control theory, and treats the anharmonic
term as a perturbation.
\subsection{Inverse engineering and perturbation theory}
In this section the quartic term $V_1= - \eta[x-x_0(t)]^4$ in (\ref{H+V}) is considered as a perturbation.
From the first-order perturbation theory, the wave function that evolves with (\ref{H+V}) may be approximated as
$$
|\widetilde{\psi}(t_f) \ra \simeq |\psi_n(t_f)\ra -\frac{i}{\hbar}\int_0^{t_f}\!\! dt\,
U_0(t_f,t)V_1(t)|\psi_n(t)\ra,
$$
where $U_0$ is the evolution operator for the Hamiltonian (\ref{Hamiltonian}).
We are interested in the time-averaged anharmonic energy
\beq
\label{average perturbative energy}
\overline{V}_1 = \frac{1}{t_f}\int_{0}^{t_f} \langle\psi_n (t)|V_1(t)|\psi_n (t)\rangle dt.
\eeq
Our goal is to minimize it, so that trajectories
calculated for the harmonic trap remain useful.
A lengthy but straightforward calculation gives
\beqa
\overline{V}_1 = [6n(n+1)+3] \eta \left(\frac{\hbar}{2 m \omega_0}\right)^2 + \frac{\eta}{t_f} \int_{0}^{t_f} \left[\left(\frac{\ddot{x}_c}{\omega^2_0} \right)^4 + \frac{3(2n+1) \hbar}{m \omega_0} \left( \frac{\ddot{x}_c}{\omega^2_0} \right)^2\right] dt.
\eeqa
When the condition
\beq
\label{conditiontf}
t_f \ll \frac{1}{\omega_0} \sqrt[4]{\frac{m d^2\omega_0}{3( 2n+1)\hbar}}
\eeq
is satisfied (i.e. $t_f \ll 400$ ms for the parameters considered in this paper), then $(\ddot{x}_c/ \omega^2_0)^4 \gg (\hbar/m\omega_0)(\ddot{x}_c/ \omega^2_0)^2$, so that the time-averaged perturbative energy can be further simplified as
\beq
\label{reduced}
\overline{V}_1 \simeq [6n(n+1)+3] \eta \left(\frac{\hbar}{2 m \omega_0}\right)^2 +\frac{\eta}{t_f} \int_{0}^{t_f} \left( \frac{\ddot{x}_c}{\omega^2_0} \right)^4 dt,
\eeq
where the first term is constant, and the second one depends on the trajectory $x_c$. In the following we shall minimize the second term in (\ref{reduced}) using OCT. In all examples $n=0$.
\section{Optimal Control Theory}
In this section, we set the optimal control problem and define the state variables and (scalar) control function,
\beqa
x_1 = x_c,~ x_2 = \dot{x}_c, ~u(t) = x_c -x_0,
\eeqa
such that (\ref{classical}) gives a system of equation, $\dot{\textbf{x}} = \textbf{f}(\textbf{x}(t), u)$, that is,
\beqa
\label{system-1}
\dot{x}_1  &=&  x_2,
\\
\label{system-2}
\dot{x}_2 &=& - \omega^2_0 u.
\eeqa
Our optimal control problem is to minimize the cost function, see (\ref{reduced}) and (\ref{classical}),
\beq
\label{cost}
J = \int^{t_f}_0 u^4 dt.
\eeq
We may in addition set a bound for the displacement between the center of the mass of cold atoms and the trap center, i.e. $|u (t)| \leq \delta$ ($\delta >0$),
so that the instantaneous transient energy is never too high.
The boundary conditions (\ref{con0}) and (\ref{contf}) imply that
the dynamical system starts at $\{x_1(0)=0, x_2 (0)= 0\}$, and ends up at $\{x_1(t_f)=d, x_2 (t_f)= 0\}$ for some fixed bound $\delta$,
with $u(0)=0$ and $u(t_f) =0$. The boundary conditions, $u(t)=0$ for $t\leq 0$ and $t\geq t_f$, guarantee that the center of mass
and the trap center coincide before and after the transport, which implies that appropriate jumps at these points are required for the optimal control
to match the boundary conditions, without affecting the cost.
To minimize the cost function (\ref{cost}), we apply Pontryagin's maximal principle \cite{LSP}.
The control Hamiltonian is
\beq
\label{controlH}
H_c = -p_0 u^4 + p_1 x_2 - p_2 u,
\eeq
where $p_0$ is a normalization constant, and $p_1$, $p_2$ are Lagrange multipliers. Pontryagin's maximal principle states that
for the dynamical system $\dot{\textbf{x}} = \textbf{f}(\textbf{x}(t), u)$, the coordinates of the extremal vector $\textbf{x} (t)$ and of
the corresponding adjoint sate $\textbf{p} (t)$ formed by Lagrange multipliers fulfill
$ \dot{\textbf{x}} = \partial H_c/\partial \textbf{p}$ and $\dot{\textbf{p}} = - \partial H_c/\partial \textbf{x}$,
which gives the two costate equations
\beqa
\label{costate1}
\dot{p}_1 &=& 0,
\\
\label{costate2}
\dot{p}_2 &=& -p_1,
\eeqa
such that for almost all $0 \leq t \leq t_f$, the values of the control maximize $H_c$, and
$H_c [\textbf{p}(t),\textbf{x}(t),u(t)] =c$, with $c$ being a positive constant.
\subsection{Unbounded control}
\begin{figure}[]
\begin{center}
\scalebox{0.7}[0.7]{\includegraphics{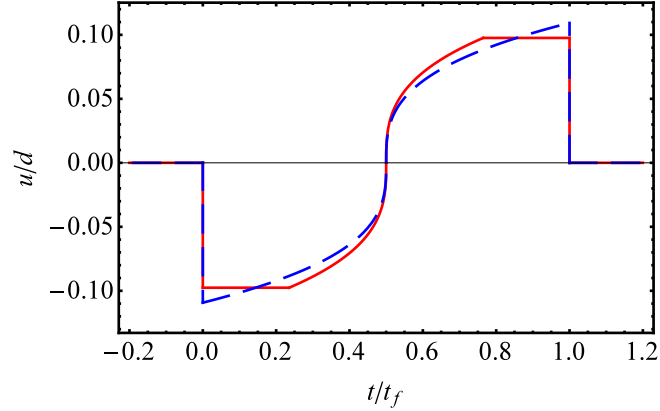}}
\caption{\label{u} (Color online) Control function for the optimization with unbounded (dashed blue line) and bounded controls (solid red line) with the bound $\delta=0.89 \delta_0$, where $\delta_0=14d/(3 \omega^2_0 t^2_f )$.
Parameters: $d=1 \times 10^{-2}$~m, $\omega_0=2 \pi\times 20 $~Hz, and $t_f=0.052$~s.}
\end{center}
\end{figure}

According to the maximum principle, the control $u(t)$ maximizes the control Hamiltonian at each time.
For simplicity, we choose $p_0=1/4$, the control Hamiltonian (\ref{controlH}) becomes $H_c=- u^4/4 + p_1 x_2 - p_2 u$, so that
$\partial H_c/ \partial u =0$ gives the function of $u(t)$, $-u^3 -p_2 u =0$, which maximizes the control Hamiltonian.
The solution can be written as $u(t) =-\sqrt[3]{p_2} = -{(-c_1 t + c_2)}^{1/3}$,
when substituting the solutions, $p_1 = c_1$, $p_2 = -c_1 t + c_2$,
calculated from (\ref{costate1}) and (\ref{costate2}), with the constants $c_1$ and $c_2$.
Solving the system of differential equations (\ref{system-1}) and (\ref{system-2}), and applying
the boundary conditions $x_1(0)=0$, $x_1(t_f) =d$, and $x_2(0)=x_2(t_f)=0$,
we get the control function, see Fig. \ref{u},
\beq
u(t) = \frac{14 d}{3 \omega^2_0  t^2_f} \left[2 \left(\frac{t}{t_f}\right)-1 \right]^{\frac{1}{3}},
\eeq
with $c_1=5488d^3/27 \omega_0^6  t_f^7$ and $c_2=2744d^3/27 \omega_0^6 t_f^6$, and the classical trajectory
\beq
\label{quasioptimal}
x_c (t) =\frac{3d}{8} \left[1-2  \left(\frac{t}{t_f}\right) \right]^{\frac{7}{3}}+\frac{7d}{4} \left(\frac{t}{t_f}\right)-\frac{3d}{8}.
\eeq
The trajectory (\ref{quasioptimal}), see Fig. \ref{figtrajectory} (a), is consistent with the result calculated from the Euler-Lagrange equation,
see Appendix A. Since the trajectory does not satisfy the boundary conditions $\dot{x}_2 (0)=\dot{x}_2 (t_f)=0$,
this is a ``quasi-optimal" trajectory.

To guarantee $u(t)=0$ at $t \leq 0$ and $t \geq t_f$ and match the boundary conditions, the control function $u(t)$ in unbounded control
has to be complemented by the appropriate jumps, see Fig. \ref{u},
\beqa
\label{control function-unbounded}
u (t) = \left\{\begin{array}{lll}
0, & t \leq 0
\\
\frac{14 d}{3 \omega^2_0  t^2_f} \left[2 \left(\frac{t}{t_f}\right)-1 \right]^{\frac{1}{3}}, & 0<t <t_f
\\
0, & t \geq t_f
\end{array}\right..
\eeqa
From (\ref{classical}), the trap trajectory $x_0(t)$ is thus calculated as $x_0= x_c -u(t)$, see Fig. \ref{figtrajectory} (b).
Since the control function $u(t)$ in unbounded control is discontinuous, the trap is allowed to
move suddenly at $t=0$ and $t=t_f$.
\begin{figure}[]
\begin{center}
\scalebox{0.7}[0.7]{\includegraphics{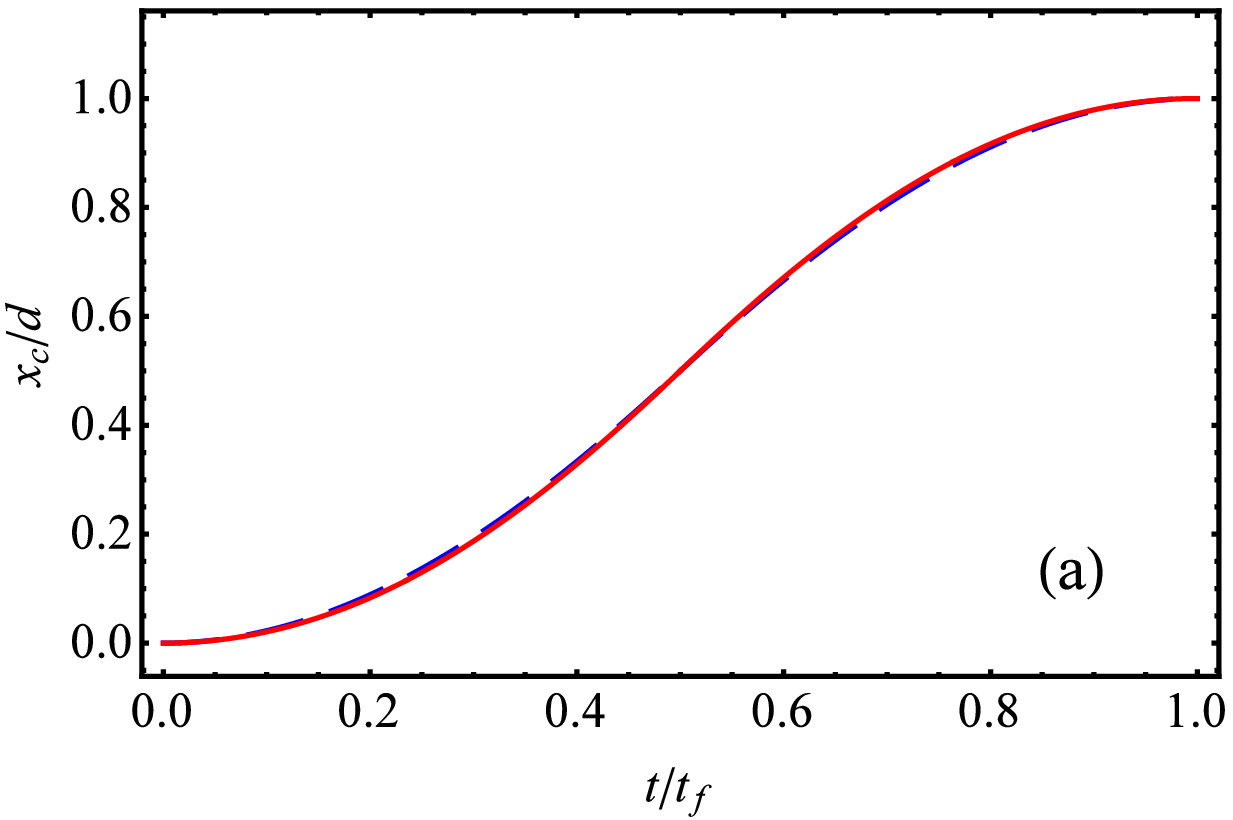}}
\\
\scalebox{0.7}[0.7]{\includegraphics{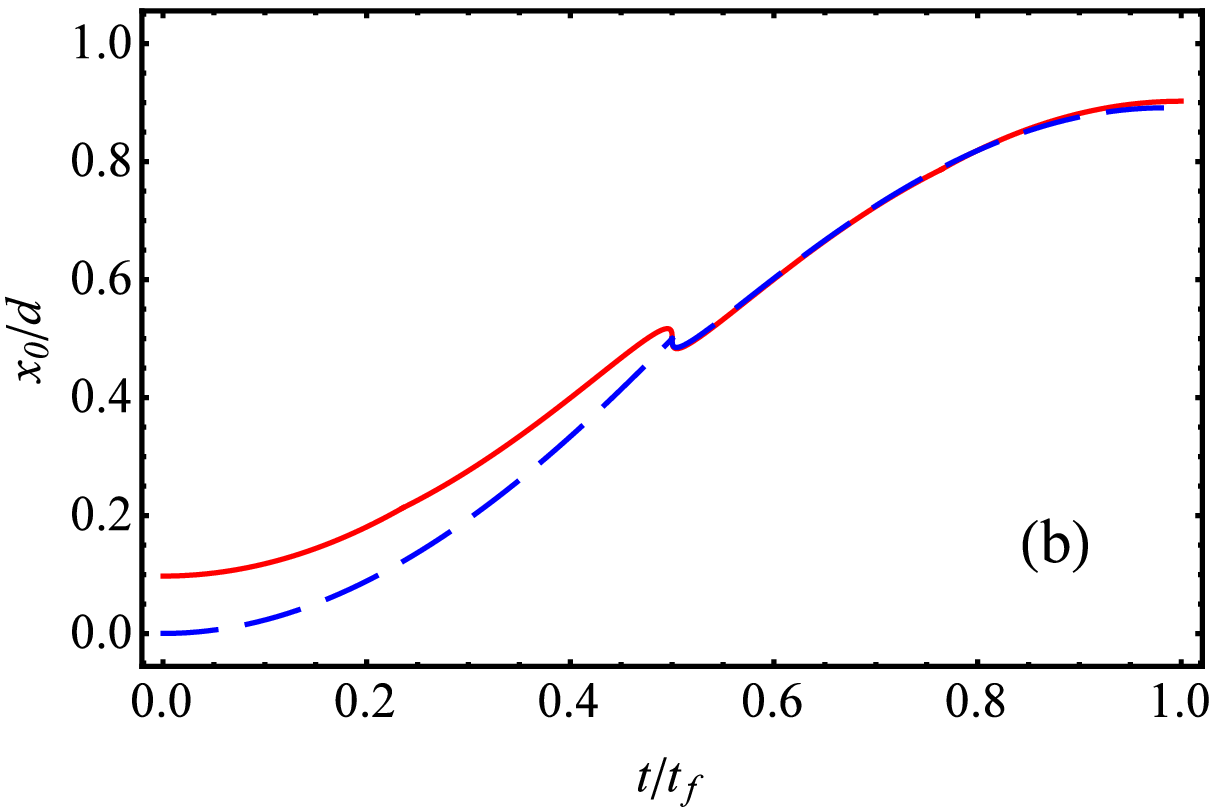}}
\caption{\label{figtrajectory} (Color online) (a) Optimal trajectories of $x_c$, the center of transport modes, for unbounded (dashed blue line) and bounded controls (solid red line); (b) Optimal trajectories of $x_0$, the trap center, for unbounded (dashed blue line) and bounded controls (solid red line). Same parameters as Fig.~\ref{u}. }
\end{center}
\end{figure}
\subsection{Bounded control}
For the bounded control, we set $|u(t)| \leq \delta$. Therefore, we
assume that the control function, see Fig. \ref{u}, is
\beqa
\label{control function-energy}
u (t) = \left\{\begin{array}{lll}
0, & t \leq 0
\\
- \delta, & 0<t <t_1
\\
-{(-c_1 t + c_2)}^{1/3}, & t_1<t<t_1+ t_2
\\
\delta, & t_1 + t_2 < t <t_f
\\
0, & t \geq t_f
\end{array}\right.,
\eeqa
where, because of $t_f = 2 t_1 + t_2$ due to the symmetry, $c_2= c_1 t_f/2$, the two switching times $t_1$
and $t_2$ are given by $t_1= t_f/2-\delta ^3/c_1$, $t_2= 2 \delta^3 /c_1$.
Substituting (\ref{control function-energy}) into (\ref{system-2}), we have
\beqa
\dot{x}_c(t) = \left\{
\begin{array}{lll}
{\omega_{0}^{2} \delta t},& 0\leq t<t_1\\
-\frac{3}{4}\omega_{0}^{2}{c_1}^{\frac{1}{3}}( t-\frac{t_f}{2})^{\frac{4}{3}} +c_3, & t_1<t<t_1+t_2
\\
-{\omega_{0}^2}\delta(t-t_f) ,& t_1+t_2<t\leq t_f
\end{array}
\right.,
\label{deriv}
\eeqa
which finally gives
\beqa
\label{optimal}
x_c(t) = \left\{
\begin{array}{lll}
\frac{1}{2}\omega_{0}^{2} \delta t^2, & 0 \leq t<t_1
\\
- \frac{9\omega^2_0}{28} {c_1}^{\frac{1}{3}}(t-\frac{t_f}{2})^{\frac{7}{3}} +c_3 t+c_4, & t_1<t<t_1+t_2
\\
d-\frac{1}{2}{\omega_{0}^2} \delta(t-t_f)^2,& t_1 +t_2 < t \leq t_f
\end{array}
\right..
\eeqa
The continuity of velocity $\dot{x}_c$ at $t=t_1$ and $t=t_1+t_2$ determines
$$
c_3 =  \frac{1}{2} \omega^2_0 \delta t_f-\frac{1}{4c_1}\omega^2_0 \delta^4.
$$
%
Furthermore, the continuity of  $x_c$ at $t=t_1$ determines
$$
c_4= -\frac{1}{3}\omega^2_0 t^2_f \delta + \frac{1}{8 c_1} \omega^2_0 t_f \delta^4 -\frac{1}{14 c^2_1}\omega^2_0 \delta^7.
$$
The constants $c_2$, $c_3$, $c_4$ and two switching times $t_1$ and $t_2$ are all dependent of $c_1$, and can be found
from the continuity of the trajectory $x_c$ at $t=t_1+t_2$,
\beq
\label{c1}
c_1=2\omega_0 \sqrt{\frac{\delta^7}{7({\omega_0}^2{ t^2_f} \delta-4d)}}.
\eeq
Figure \ref{figtrajectory} shows the trajectories of the center of mass and trap center, $x_c (t)$ and $x_0 (t)$. Due to the discontinuity of
the control function $u(t)$ at $t=0$ and $t_f$, the trajectory of the trap center $x_0$ at the edges is not continuous,
but the classical trajectory $x_c(t)$ satisfies the boundary conditions, $x_c(0)=0$ and $x_c(t_f)=d$.
From (\ref{c1}), we see that the bound should satisfy
\beq
\delta \geq \frac{4d}{\omega^2_0 t^2_f}
\eeq
to make $c_1$ real.
This gives the minimal possible time $t_f^{min} = (2/\omega_0)\sqrt{d/\delta}$ for a given bound $\delta$ \cite{Xi12}.
In addition, we get
\beq
\delta = \delta_0 = \frac{14d}{3 \omega^2_0 t^2_f}
\eeq
to make $t_1=0$, which implies that for $\delta_0$ the bounded control tends to the unbounded one.
Combining these results $\delta$ is restricted to the interval
\beq
\frac{14d}{3 \omega^2_0 t^2_f} \geq \delta \geq \frac{4d}{\omega^2_0 t^2_f}
\eeq
for a non-trivial bounded control.
\begin{figure}[tbp]
\begin{center}
\scalebox{0.58}[0.6]{\includegraphics{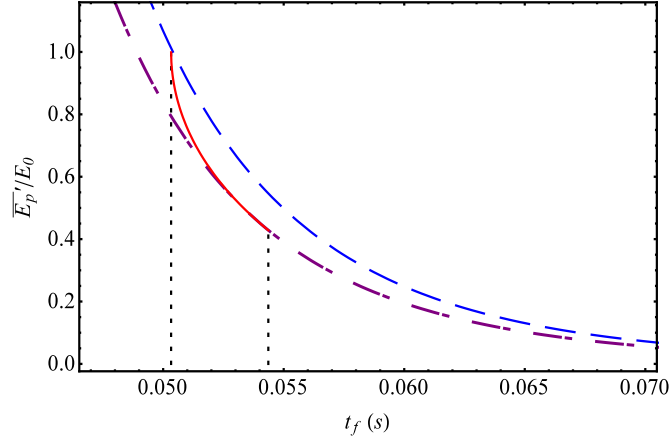}}
\caption{\label{figenergy} (Color online) Dependence of time-averaged anharmonic pertubative energy $\overline{E'_p}$ (in units of $E_0 = \eta \delta^4$) for different protocols, such as optimal trajectories -the ones that minimize the anharmonic energy- with bounded (red solid line) and unbounded (purple dot-dashed line) controls. The values for the trajectory that minimizes the harmonic potential energy (blue dashed line) are also shown for comparison.
The vertical dotted black lines mark the interval $(2/\omega_0) \sqrt{d/\delta} \leq t_f \leq (1/\omega_0) \sqrt{14 d/3 \delta}$. Parameters:
$\delta=0.89\delta_0$, and $\delta_0 = 14d/(3 \omega^2_0 t^2_f)$, $d=1 \times 10^{-2}$~m, $\omega_0=2 \pi\times 20 $~Hz, $w_0=50\lambda$, $\lambda=1060$~nm, and $m= 1.44269 \times 10^{-25}$~kg the mass of $^{87}\mbox{Rb}$ atoms.
}
\end{center}
\end{figure}
\section{Time-averaged anharmonic energy}
To analyze the effect of the optimization, we define the time-averaged anharmonic energy as
\beq
\overline{E'_p}\equiv \frac{1}{t_f} \int_0^{t_f} E'_p dt = \frac{1}{t_f} \int_0^{t_f} \eta (x_c-x_0)^4 dt.
\eeq
Using the optimal trajectory (\ref{optimal}) with the bounded control (\ref{control function-energy}), we obtain, see Fig. \ref{figenergy} (b),
\beq
\label{energy-e}
\overline{E'_p}=\eta \delta^4 \left(1-\frac{4\sqrt{7}}{7}\sqrt{1-\frac{4d}{\omega^2_0 {t^2_f} \delta}} \right),
\eeq
which takes the minimal value
\beq
\label{energy bound}
\overline{E'_p}_{min}=\frac{392 \eta d^4 }{9 \omega^8_0 t^8_f},
\eeq
when $\delta= 14 d/3\omega_0^2t_f^2$. This minimal value for anharmonic potential energy is also the exact expression for optimal unbounded control.
When $t_f = t_f^{min} = (2/\omega_0)\sqrt{d/\delta}$, (\ref{energy-e}) also gives the maximum value, $E_0= \eta \delta^4$,
for anharmonic potential energy with the bounded control. The time-averaged anharmonic perturbative energy, $\overline{E'_p}$, depends on $t^{-8}_f$.
The scaling law found here is quite different from the one for trap expansions \cite{Chenenergy10},
which is $\overline{E_n} \propto t^{-2}_f$.
Figure \ref{figenergy} compares the time-averaged anharmonic energy for bounded and unbounded optimal trajectories.

\begin{figure}[]
\begin{center}
\scalebox{0.59}[0.6]{\includegraphics{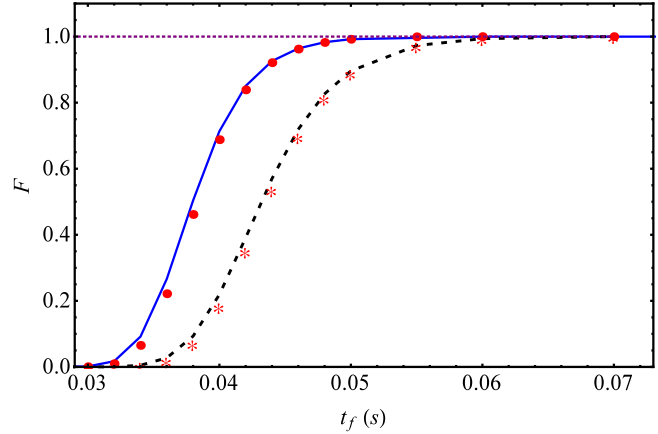}}
\caption{\label{figfidelity} (Color online) Fidelity for Gaussian trap versus $t_f$ for different protocols, including
polynomial ansatz (dashed black line) and ``quasi-optimal" solution (blue solid line). The dotted purple line represents
the perfect transport for the (unperturbed) harmonic trap for comparison. The fidelities for a trap that includes only
quadratic and quartic terms (with ``$\bullet$"  and ``$\ast$") are indistinguishable from the ones for the Gaussian.
Same parameters as Fig. \ref{figenergy}. }
\end{center}
\end{figure}

If the perturbative energy is constrained by some maximally allowed value $\overline{E'_M}$, $t_f$ should satisfy, see (\ref{energy bound}),
\beq
t_f \geq \frac{1}{\omega_0} \left(\frac{392 \eta d^4}{9 \overline{E'_M}}\right)^{1/8}.
\eeq
This is different from the minimal time discussed before $t^{min}_f = (2/\omega_0)\sqrt{d/\delta}$ \cite{Xi12}
as different constraints are imposed.

As a matter of fact, different physical constraints require different optimal trajectories for atomic transport. Other ``quasi-optimal" trajectory, $x_c(t) = d (t/t_f)^2 (3-2 t/t_f)$, minimizes the time-averaged
(harmonic) potential energy,
\beq
\overline{E_p}\equiv \frac{1}{t_f} \int_0^{t_f} E_p dt=\frac{1}{t_f} \int_0^{t_f} \frac{1}{2}m\omega_0^2 (x_c-x_0)^2 dt,
\eeq
which gives \cite{Xi12}
\beq
\overline{E_{p}}_{min}=\frac{6m d^2}{\omega^2_0 {t^4_f}}.
\eeq
However, the time-averaged anharmonic energy for such  ``quasi-optimal" trajectory is calculated as
\beq
\overline{E'_p}= \frac{1296 \eta d^4}{5 \omega^8_0 t^8_f},
\eeq
which is larger than the minimal value $\overline{E'_p}_{min}$ in (\ref{energy bound}), see Fig. \ref{figenergy}.

Finally, to see the effect of the anharmonic energy minimization on the fidelity of the final state with respect to the
one for purely harmonic transport,
$F = |\la  \psi_0 (t_f) | \tilde{\psi} (t_f)\ra|$, the final state $\tilde{\psi}(t_f)$ is calculated by solving the time-dependent Schr\"{o}dinger equation numerically with the split-operator
method. Fig. \ref{figfidelity} shows that the optimal trajectory gives a fidelity of nearly one except for very short times.
We have also computed the fidelity for a Gaussian potential with the same harmonic and quadratic terms \cite{Weiss,Book}.
This gives results which are indistinguishable from the quartic model.

%
\section{Conclusion}
We have found optimal shortcut protocols for fast atomic transport in anharmonic traps. We combine invariant-based inverse engineering, perturbation theory, and
optimal control theory to minimize the contribution
of the anharmonicity to the potential energy. Numerical calculation of the fidelity demonstrates that the
designed optimal trajectory can provide fast and faithful transport in a Gaussian trap.
These results can be readily extended to other anharmonic traps like the power-law trap \cite{Qi2015}, or to
the transport of Bose-Einstein Condensates \cite{Erikcond12}.

\section*{Acknowledgments}
This work was partially supported by the NSFC (11474193),
the Shuguang Program (14SG35), the Specialized Research Fund for the Doctoral Program (2013310811003),
the Program for Eastern Scholar,  the grant NEXT ANR-10-LABX-0037 in the framework of the Programme des Investissements
d'Avenir, the Basque Government (Grant IT472-10), MINECO (Grant FIS2015-67161-P), and the program UFI 11/55 of
UPV/EHU.

\appendix

\section{Euler-Lagrange Equation}
Here we use the Euler-Lagrange equation to minimize
$
\int_0^{t_f} (\ddot{x}_c /\omega_0^2)^4 dt.
$
We set $\mathcal{L}(t,x_c,\dot{x}_c,\ddot{x}_c) =  (\ddot{x}_c /\omega_0^2)^4 $, so that the Euler-Lagrange equation,
\beq
\label{euler-lagrange equation}
\frac{\partial \mathcal{L}}{\partial x_c}-\frac{d}{dt}\left( \frac{\partial \mathcal{L}}{\partial \dot{x}_c}\right)+\frac{d^2}{dt^2}\left( \frac{\partial \mathcal{L}}{\partial \ddot{x}_c}\right)=0,
\eeq
gives
\beq
\frac{d^2}{dt^2}\left(\ddot{x}_c\right)^3=0.
\eeq
The solution for $x_c(0)=0$, $x_c(t_f)=d$, and $\dot{x}_c(0)=\dot{x}_c(t_f)=0$
is
\beq
x_c(t)=\frac{3d}{8} \left[1-2\left(\frac{t}{t_f}\right) \right]^{\frac{7}{3}}+\frac{7d}{4}\left(\frac{t}{t_f}\right)-\frac{3d}{8}.
\eeq


\section*{References}

\begin{thebibliography}{99}
%
\bibitem{HanschNature2001}H\"ansel W, Hommelhoff P,  H\"ansch T W and Reichel J 2001 \textit{Nature} \textbf{413} 498

\bibitem{HanschPRL2001}H\"ansel W, Reichel J, Hommelhoff P and H\"ansch T W 2001 \textit{Phys. Rev. Lett.} \textbf{86} 608

\bibitem{Ketterle2002}Gustavson T L, Chikkatur A P, Leanhardt A E, G\"orlitz A, Gupta S, Pritchard D E and  Ketterle W 2001 \textit{Phys. Rev. Lett.} \textbf{88} 020401

\bibitem{Leibfried}Rowe M A, Ben-Kish A, DeMarco B, Leibfried D, Meyer V, Beall J, Britton J,  Hughes J,  Itano W M,  Jelenkovic B, Langer C, Rosenband T and Wineland D J 2002 \textit{Quant. Inf. Comp.} \textbf{4} 257

\bibitem{ions}Reichle R, Leibfried D, Blakestad R B, Britton J, Jost J D, Knill E, Langer C, Ozeri R, Seidelin S and Wineland D J 2006 \textit{Fortschr. Phys.} \textbf{54} 666

\bibitem{Bowler}Bowler R, Gaebler J, Lin Y, Tan T R, Hanneke D, Jost J D, Home J P, Leibfried D and Wineland D J 2012 \textit{Phys. Rev. Lett.} \textbf{109} 080502

\bibitem{Walther}Walther A, Ziesel F, Ruster T, Dawkins S T, Ott K, Hettrich M, Singer K, Schmidt-Kaler F and Poschinger U 2012 \textit{Phys. Rev. Lett.} \textbf{109} 080501


\bibitem{David08}Couvert A, Kawalec T, Reinaudi G and Gu\'ery-Odelin D 2008 \EPL \textbf{83} 13001

\bibitem{Calarco09}Murphy M, Jiang L, N. Khaneja and Calarco T 2009 \textit{Phys. Rev. A} \textbf{79} 020301(R)

\bibitem{TransAPL10}Chen D, Zhang H, Xu X, Li T and Wang Y Z 2010 \textit{Appl. Phys. Lett.} \textbf{96} 134103

\bibitem{Masuda10}Masuda S and Nakamura K 2010 \textit{Proc. R. Soc. A} \textbf{466} 1135

\bibitem{Erik11}Torrontegui E, Ibanez S, Chen X, Ruschhaupt A, Gu\'ery-Odelin D and Muga J G 2011 \textit{Phys. Rev. A} \textbf{83} 013415

\bibitem{Xi12}Chen X, Torrontegui E, Stefanatos D, Li J S and Muga J G 2011 \textit{Phys. Rev. A} \textbf{84} 043415

\bibitem{Erikcond12}Torrontegui E, Chen X, Modugno M, Schmidt S, Ruschhaupt A and Muga J G 2012 \NJP \textbf{14} 013031

\bibitem{Mikel13}Palmero M, Torrontegui E, Gu\'{e}ry-Odelin D and Muga J G 2013 \textit{Phys. Rev. A} \textbf{88} 053423

\bibitem{Uli14}F\"urst H. A, Goerz M H, Poschinger U G, Murphy M, Montangero S, Calarco T, Schmidt-Kaler F, Singer K and Koch C P 2014 \NJP \textbf{16} 075007

\bibitem{CampoPRX}Deffner S, Jarzynski C and Del Campo A 2014 \textit{Phys. Rev. X} \textbf{4} 021013

\bibitem{Lu14}Lu X J, Muga J G, Chen X, Poschinger U G, Schmidt-Kaler F and Ruschhaupt A 2014 \textit{Phys. Rev. A} \textbf{89} 063414

\bibitem{Mikel14}Palmero M, Bowler R, Gaebler J P, Leibfried D and Muga J G 2014 \textit{Phys. Rev. A} \textbf{90} 053408

\bibitem{David14}Gu\'{e}ry-Odelin D and Muga J G 2014 \textit{Phys. Rev. A} \textbf{90} 063425

\bibitem{Qi2015}Zhang Q, Chen X and Gu\'{e}ry-Odelin D 2015 \textit{Phys. Rev. A} \textbf{92} 043410

\bibitem{Li2010}Stefanatos D, Ruths J and Li J S 2010 \textit{Phys. Rev. A} \textbf{82} 063422
%
\bibitem{Ricci}Bertoldi A and Ricci L 2010 \textit{Phys. Rev. A} \textbf{81} 063415

\bibitem{AlonsoNJP}Alonso J, Leupold F M, Keitch B C and Home J P 2013 \NJP \textbf{15} 023001

\bibitem{OptTrap} Ahmadi P, Behinaein G, Timmons B P and Summy G S 2006 \JPB \textbf{39} 1159

\bibitem{Chenenergy10}Chen X and Muga J G 2010 \textit{Phys. Rev. A} \textbf{82} 053403

\bibitem{Erik3DPRA} Torrontegui E, Chen X, Modugno M, Ruschhaupt A, Gu\'{e}ry-Odelin D and Muga J G 2012 \textit{Phys. Rev. A} \textbf{85} 033605

\bibitem{LuPRA2014}Lu X J, Chen X, Alonso J and Muga J G 2014 \textit{Phys. Rev. A} \textbf{89} 023627

\bibitem{NJP14}L\'{e}onard J, Lee M, Morales A, Karg T M, Esslinger T and Donner T 2014 \NJP \textbf{16} 093028

\bibitem{Weiss} Kinoshita T, Wenger T and Weiss D S 2004 \textit{Science} \textbf{305} 1125

\bibitem{LR}Lewis H R and Riesenfeld W B 1969 \textit{J. Math. Phys.} \textbf{10} 1458

\bibitem{Leach} Lewis H R and Leach P G 1982 \textit{J. Math. Phys.} \textbf{23} 2371

\bibitem{PRL104} Chen X, Ruschhaupt A, Schmidt S, Del Campo A, Gu\'{e}ry-Odelin D and Muga J G 2010 \textit{Phys. Rev. Lett.} \textbf{104} 063002

\bibitem{saraPRL104} Ib\'{a}\~{n}ez S, Chen X, Torrontegui E, Muga J G and Ruschhaupt A 2012 \textit{Phys. Rev. Lett.} \textbf{109} 100403

\bibitem{Berry} Berry M V 2009 \textit{J. Phys. A} \textbf{42} 365303

\bibitem{ChenPRA11} Chen X, Torrontegui E and Muga J G 2011 \textit{Phys. Rev. A} \textbf{83} 062116

\bibitem{LSP} Pontryagin L S {\it et al., The Mathematical Theory of Optimal Processes} (Interscience Publishers, New York) 1962

\bibitem{Book} Saleh B E A and Teich M C \textit{Fundamentals of Photonics}, 2nd ed. (Wiley, New York) 2007



\end{thebibliography}
\end{document}